\documentclass{article}

\usepackage{microtype}
\usepackage{graphicx}
\usepackage{subcaption}
\usepackage{booktabs} 

\usepackage{hyperref}

\usepackage[preprint]{icml2026}

\usepackage{amsmath}
\usepackage{amssymb}
\usepackage{mathtools}
\usepackage{amsthm}

\usepackage[capitalize,noabbrev]{cleveref}

\theoremstyle{plain}

\theoremstyle{definition}

\theoremstyle{remark}

\usepackage[textsize=tiny]{todonotes}

\begin{document}

\twocolumn[
  \icmltitle{Position: No Retroactive Cure for Infringement during Training}



  \icmlsetsymbol{equal}{}

  \begin{icmlauthorlist}
    \icmlauthor{Satoru Utsunomiya}{utokyo}
    \icmlauthor{Masaru Isonuma}{utokyo,nii,tohoku,riken}
    \icmlauthor{Junichiro Mori}{utokyo}
    \icmlauthor{Ichiro Sakata}{utokyo}
  \end{icmlauthorlist}

  \icmlaffiliation{utokyo}{The University of Tokyo, Tokyo, Japan}


  \icmlaffiliation{nii}{NII LLMC, Tokyo, Japan}
  \icmlaffiliation{tohoku}{Tohoku University, Miyagi, Japan}
  \icmlaffiliation{riken}{RIKEN, Tokyo, Japan}

  \icmlcorrespondingauthor{Satoru Utsunomiya}{utsunomiya-satoru200229@g.ecc.u-tokyo.ac.jp}

  \icmlkeywords{Machine Learning, ICML}

  \vskip 0.3in
]



\printAffiliationsAndNotice{}  

\begin{abstract}
As generative AI faces intensifying legal challenges, the machine learning community has increasingly relied on \emph{post-hoc mitigation}---especially machine unlearning and inference-time guardrails---to argue for compliance. \textbf{This paper argues that such post-hoc mitigation methods cannot retroactively cure liability from unlawful acquisition and training, because compliance hinges on data lineage, not the outputs.}
Our argument has three parts. First, unauthorized copying/ingestion can be a legally complete \emph{completed act}, and model weights may operate as \emph{fixed copies} that retain training-derived expressive value, making later filtering beside the point for infringement. Second, \emph{contract} and \emph{tort/unfair-competition} rules---via licenses, terms of service, and anti-free-riding principles---can independently restrict access and use, often bypassing copyright defenses (e.g., fair use or TDM exceptions). Third, since value from protected inputs can persist in weights, remedies such as \emph{unjust enrichment} and \emph{disgorgement} may require stripping gains and, in some cases, reaching the model itself. We therefore argue for a shift from \emph{Post-Hoc Sanitization} to verifiable \emph{Ex-Ante Process Compliance}.
\end{abstract}

\section{Introduction}

The rapid rise of modern foundation models has been enabled by what we might call the data ingestion paradigm: training on unprecedented volumes of text and images drawn from across the open web~\cite{brown2020language, bommasani2021opportunities}. But since these datasets are so large, it has become technically and economically unrealistic to identify and remove every instance of harmful material or protected content before training (i.e., ex-ante filtering). As a result, many developers have proceeded with large-scale training on broadly collected data and have often defended this practice under the banner of fair use.

This approach has triggered serious legal challenges that now threaten the foundations of the AI development ecosystem. Allegations include copyright infringement, privacy violations, and breaches of licensing terms. Rights holders have responded forcefully, arguing that training on protected works and sensitive personal data without authorization is unlawful. High-profile lawsuits—including \textit{The New York Times v. OpenAI}~\cite{nytopenai} and \textit{Getty Images v. Stability AI}~\cite{gettystability}—have brought these issues into sharp focus. Importantly, these disputes are not limited to whether model outputs resemble protected works. They also raise broader questions about legal liability for the acts of copying, storing, and using protected or personal data during the training process itself. In this sense, the litigation risk concerns not merely product behavior, but the viability of model development as such.

Against this backdrop—where ex-ante filtering is infeasible and legal pressure is increasing—the research community and industry have increasingly turned toward post-hoc mitigation strategies, namely interventions applied after a model has already been trained. For instance, some generative AIs have introduced opt-out mechanisms intended to prevent the generation of specific copyrighted content, and academic interest in machine unlearning has grown as a technique for removing the influence of particular training data from an already-trained model. These approaches are often presented as pragmatic ways to reduce legal exposure. They are supported by an engineering intuition we call output equivalence: if the system no longer produces infringing or harmful outputs, then it is effectively equivalent to a model that never learned from the problematic data—and therefore should be treated as compliant.

\textbf{This paper argues that such post-hoc mitigation strategies do not, as a matter of law, eliminate legal liability arising from the training process.} Across jurisdictions, developers can face multiple, overlapping forms of exposure based on how training data are acquired and used, including \emph{copyright infringement}, \emph{contract-based claims} (e.g., license or terms-of-service violations), and \emph{tort/unfair competition theories}. Because these liabilities can be established based on acts of acquisition, copying, and reproduction---not solely on a model's current output behavior---adjusting outputs after training generally cannot retroactively cure an unlawful training-stage act.

As a result, incurring legal liability can impose substantial costs on developers, potentially resulting in severe remedies such as algorithmic disgorgement and significant monetary exposure. This is why the AI industry must shift away from treating post-hoc mitigation as a compliance strategy and toward verifiable \emph{ex-ante process compliance}.

In the remainder of this paper, we proceed from context-setting to premises, then to doctrinal analysis, and finally to practical implications. Section~2 surveys related work at the intersection of engineering and law, emphasizing a persistent disconnect between them. Section~3 states and motivates our legal premise by tying unauthorized dataset creation and training to the reproduction right across jurisdictions and showing why key defenses (fair use/TDM) are increasingly constrained. Building on this premise, Section~4 presents our core copyright analysis—introducing the completed-act framing and examining model weights through fixation and perceptibility—showing why suppressing outputs does not retroactively cure training-stage exposure. We then broaden the lens in Section~5 to contract and tort/unfair competition theories that can independently constrain data acquisition and reuse, and in Section~6 to remedial doctrines such as unjust enrichment and disgorgement. Section~7 addresses leading counterarguments and clarifies the limits of defenses commonly invoked in practice. Finally, Section~8 translates our findings into concrete directions for model development and governance.

\section{Related Work: The Disconnect Between Engineering and Law}

Research on legal compliance for generative AI is marked by a clear divide between disciplines. In engineering, most works evaluate technical interventions using engineering metrics while taking legal validity for granted. In legal scholarship, by contrast, much of the discussion focuses on abstract doctrines and policy, often without assessing whether the emerging technical remedies, such as post-hoc mitigation techniques, are legally meaningful. This section highlights this epistemic gap and our contributions.

\subsection{Engineering Mitigation and the Output Equivalence Assumption}

The engineering literature has proposed a wide range of techniques intended to suppress or remove undesirable information from trained models. Machine unlearning methods—including SISA~\cite{bourtoule2021machine} and gradient-based approaches~\cite{jang2023knowledge, eldan2023who}—seek to eliminate the influence of specific training examples. Relatedly, model editing techniques such as ROME~\cite{meng2022locatingeditingfactualassociations} aim to directly modify particular factual associations in model parameters. Inference-time guardrails further attempt to prevent infringing outputs through retrieval-based constraints or other runtime controls~\cite{inan2023llama}.


A central feature of this engineering work is an often implicit assumption we call output equivalence: if post-hoc interventions make a model's output distribution statistically indistinguishable from that of a counterfactual model trained without the problematic data, then the system is treated as compliant. To ensure the robustness of our legal analysis, we adopt a strong technical premise that represents the best-case performance of post-hoc mitigation methods. Specifically, for the sake of argument, we assume that post-hoc methods such as \emph{inference-time guardrails} and \emph{parameter modification} can achieve their theoretical ideal, producing a modified model whose output distribution $P^*$ is statistically identical to that of a model retrained from scratch without the infringing data:
$P^*(y|x) = P(y|x; \text{retrained model})$ for any input $x$ and output $y$.

In practice, prior studies emphasize effectiveness and utility metrics (e.g., retention of general capabilities and reduction of targeted behaviors), but they typically do not examine whether even this strongest form of output-level equivalence satisfies the doctrinal requirements of copyright law, particularly with respect to liability arising from upstream acts of ingestion, copying, and reproduction during training.

\subsection{Legal Scholarship and the Technical Blind Spot}

Legal scholarship on LLMs has largely developed along two tracks: (i) debates about the legality of training and (ii) debates about whether generated content is infringing.

On the input side, scholars generally view training as a non-expressive and transformative use. In the U.S., Lemley and Casey~\cite{lemley2021} and Sag~\cite{sag2018} argue that extracting patterns rather than expressive content supports fair use. Others raise concerns about legal uncertainty: Murray~\cite{murray2025} questions the reliability of existing protections when models are trained on large-scale datasets. In the EU, Christensen~\cite{christensen2021} explains how TDM exceptions in the DSM Directive apply, while Dermawan~\cite{dermawan2024} and Löbling et al.~\cite{lobling2024} highlight problems with opt-outs and technical ambiguity. In contrast, Japan’s framework is more permissive: Quang~\cite{quang2021} note that non-consumptive training is broadly allowed, offering legal clarity for developers.

On the output side, scholars examine when LLM-generated content might trigger infringement. Lemley~\cite{lemley2023} and Sobel~\cite{sobel2024} argue that traditional copyright tests—like the idea/expression dichotomy and substantial similarity—struggle to address the nature of generative outputs. In the EU, Rosati~\cite{rosati2024} and Dornis and Lucchi~\cite{dornis2025} warn that outputs may qualify as unauthorized derivative works, particularly when mimicking style or reproducing prompts. Zhang~\cite{zhang2025} and Murray~\cite{murray2023} suggest shifting focus from training legality to regulating output.

Although legal scholarships provide essential doctrine, it rarely asks whether post-hoc mitigation can serve as a legally cognizable cure for training-stage violations. As a result, whether clean-output models remain legally tainted by their training history remains largely unresolved. A rare explicit bridge is Marino et al.\cite{marino2025}, who frame machine unlearning as a potential regulation-aligned compliance tool despite current frictions. In contrast, we argue that for training-stage violations, unlearning and other post-hoc methods cannot, as a matter of principle, retroactively cure the infringement, since liability attaches to unlawful ingestion and copying at training time, not to downstream outputs.

\subsection{The Significance of Legal-Technical Alignment}
The primary contribution of this paper is not merely theoretical but sociotechnical. As long as the engineering and legal communities operate in isolation, the AI industry will remain exposed to various kinds of legal risks.  By bridging this epistemic gap, we provide a necessary framework for sustainable AI development, helping to prevent a future where foundational models face existential injunctions due to fundamental misconceptions about liability.

\section{Legal Baseline: Unauthorized Training as Copyright Infringement}

We adopt the following legal baseline: \textbf{When copyrighted data are used for model training without authorization, the acts of data acquisition and training themselves constitute copyright infringement, particularly by implicating the reproduction right.} We motivate this baseline by examining the definitions and recent judicial scrutiny across jurisdictions.

\subsection{Legal Framework: Why Training Is Reproduction and What the Exceptions Require}

The threshold question is whether the acts involved in AI training technically constitute reproduction under copyright law. We confirm that they do, based on statutory definitions in major jurisdictions.

\begin{itemize}
  \item \textbf{United States.}
  17~U.S.C.~§106(1) grants the exclusive right to reproduce a work in copies. A copy is defined in §101 as a fixation in a tangible medium of expression sufficiently permanent to permit it to be perceived or reproduced for more than transitory duration. Exceptions are evaluated under the fair use doctrine (§107), applying four factors: (1) purpose and character of the use, (2) nature of the copyrighted work, (3) amount and substantiality used, and (4) effect on the market.

  \item \textbf{European Union.}
  Article~2 of the InfoSoc Directive requires member states to protect against direct or indirect, temporary or permanent reproduction by any means and in any form. This broad scope includes digital copying acts. The DSM Directive (Art.~4) introduces a TDM exception, provided that the source is lawfully accessed and rights holders have not opted out through machine-readable means.

  \item \textbf{Japan.}
  Article~21 of the Copyright Act grants authors the exclusive right to reproduce (\textit{fukusei}), which is defined in Article~2(1)(xv) to include fixation in tangible form, such as server storage. Article~30-4 provides an exception for information analysis, but this is limited by a proviso excluding uses that unreasonably prejudice the interests of the copyright owner.
\end{itemize}

\subsection{Vector 1 — Corpus Building: Downloading and Storing as Reproduction}
\paragraph{Establishing the Act: Fixation via Downloading}
The act of scraping, downloading, and storing works to build a training corpus constitutes a prima facie violation of the reproduction right. In \textit{Capitol Records, LLC v. ReDigi Inc.} \cite{redigi}, the Second Circuit established that creating a new digital file on a server constitutes a reproduction since a new material object is created. The court reasoned that when a user downloads a file, the user is producing a new phonorecord. Thus, assembling a dataset like The Pile involves creating billions of unauthorized material objects.

\paragraph{Why Defenses Fail: Illicit Sources and Opt-Outs}
Recent rulings in both the US and EU demonstrate that defenses for this copying are fragile and contingent on strict conditions. In the US, the court in \textit{Bartz v. Anthropic} \cite{bartzorder} denied summary judgment regarding the use of pirated libraries (e.g., Books3), implying that the illicit provenance of data acts as a significant barrier to fair use. If the source material is stolen, the developer's lack of clean hands weighs heavily against excusing the reproduction. Similarly, in the EU, the District Court of Hamburg confirmed in \textit{Kneschke v. LAION} \cite{laionkneschke} that downloading images for datasets constitutes copyright reproduction. The judgment clarified that for TDM exceptions to apply, developers must strictly adhere to machine-readable opt-outs (e.g., robots.txt). Failure to respect these technical flags, or a lack of lawful access, renders the copying and use of the material for dataset construction infringing.

\subsection{Vector 2 — Training Ingestion: Computational Copies as Reproduction}
\paragraph{Establishing the Act: Fixation via RAM Copies}
The training process requires the repeated loading of datasets into RAM/VRAM. The legal status of these transient yet essential reproductions was addressed in \textit{MAI Systems Corp. v. Peak Computer} \cite{mai}, where the Ninth Circuit held that loading software into RAM constitutes the creation of a copy. This is because the data is rendered sufficiently permanent to be perceived, reproduced, or otherwise communicated for a period of more than transitory duration. Since AI training involves high-frequency, iterative access to these works over extended periods, these RAM copies satisfy the fixation requirement, establishing a prima facie act of reproduction.



\paragraph{Why Defenses Fail: The Collapse of Transformative Use}
The legal justification for data training may be weakened under the analysis of the first factor of fair use: the purpose and character of the use. The court’s ruling in \textit{Warhol v. Goldsmith} \cite{warholgoldsmith} establishes that a use cannot be considered transformative if it shares the commercial purpose of the original and functions as a market substitute. Generative AI models are fundamentally designed to satisfy the exact same market demand as their training data—providing consumers with expressive text, code, or imagery. Consequently, the model acts not as a tool for creating new meaning, but as a high-tech substitute that competes for the same user attention. This reality reclassifies the act of ingestion from a creative endeavor into an act of commercial usurpation. Since the ultimate purpose is to generate competing commercial value, the first factor weighs definitively against the developer, precluding a fair use defense.

\section{The Incurability of the Infringements}

Having established in Section~3 that the acts of dataset creation and training constitute copyright infringements (historical facts), this section argues that post-hoc mitigation techniques cannot retroactively cure these violations. We examine this through the lens of legal doctrines and the ontological status of model weights.

\subsection{Completed-Act Doctrine: Liability Attaches at the Moment of Copying}

Copyright infringement operates on a principle of strict liability, meaning that the legal wrong is fully consummated the instant an exclusive right is violated. It is not a fluid condition that fluctuates based on the infringer's subsequent remorse or technical patches. Legally, an infringement is a \textit{fait accompli}—an accomplished fact that exists independently of any remedial actions taken later. Once the statutory boundary is crossed, liability attaches permanently to the infringer's record. Consequently, future attempts to hide, delete, or filter the copies are legally irrelevant to the existence of the initial cause of action.

\paragraph{Dataset Creation as a Completed Fact}
The liability for unauthorized dataset creation is cemented at the exact moment of the initial unlicensed reproduction ($t=0$). Following the judicial reasoning in \textit{Bartz} \cite{bartzorder}, the mere act of unauthorized possession and reproduction constitutes an independent and complete legal injury. When a developer downloads and saves an unlicensed corpus like Books3 to their server, the statutory violation is finished at that precise second. This legal fact is incurable by subsequent disposition. The law treats this analogously to physical theft: a thief who returns stolen property may reduce their punishment, but they cannot nullify the fact that a theft occurred. Similarly, a developer cannot un-infringe a dataset simply by deleting it or applying a safety filter at a later date ($t=1$). The snapshot of liability captured at the moment of reproduction remains a permanent part of the legal reality; post-hoc filters are merely attempts to mitigate the scope of damages, not to erase the completed act of infringement.

\paragraph{Data Training as a Consummated Infringement}
The same logic of finality applies with equal force to the training phase. The infringement is perfected the moment the model's weights are updated to reflect the expressive value of the training data. In the context of generative AI, the creation of ephemeral RAM copies serves as the essential mechanism for this value extraction. Once the training process successfully converts the author's protected expression into the developer's commercial neural parameters, the usurpation of the author's labor is done and the point of no return has passed. This process effectively transmutes the right in the expression into a functional component of the model. Once this conversion is realized, the infringement becomes a historical reality that cannot be undone by any kind of updates. Subsequent efforts to unlearn specific data or filter outputs are merely post-hoc efforts to suppress the visible symptoms of the infringement; they cannot revert the legal status of the act from completed to uncommitted.

\subsection{Ongoing Copy Status: Fixation and Recoverability in Model Weights}

Under 17 U.S.C. \S 101, a work qualifies as a copy if it meets two distinct conditions: (1) fixed in a stable medium (Fixation), and (2) retrievable either directly or with the aid of a machine (Perceptibility). We analyze these separately to demonstrate that the post-mitigation model remains an infringing copy and that the post-hoc mitigation methods do not negate the underlying embodiment of the work.

\paragraph{Fixation: Weights Are Stable, Non-Transitory Files}
The first inquiry focuses on the stability of the data structure. A work is fixed if its embodiment is sufficiently permanent to exist for more than a transitory duration. In the context of generative AI, model weights constitute static files stored on physical media (e.g., hard drives). Unlike the transient signals of a live broadcast, these weights remain immutable across inference runs unless the model is retrained. Therefore, the weights satisfy the fixation requirement as persistent objects, regardless of whether output filters momentarily suppress their display.

\paragraph{Machine-Aided Perceptibility: Expression Remains Technically Recoverable}
The second inquiry concerns the technical recoverability of the expression. The statute explicitly allows copies to be perceived with the aid of a machine. This means the relevant legal test is not whether a naive user can see the work through a chat interface, but whether the work can be reconstructed by any technical means. Even if safety filters block standard outputs, the expression remains embodied in the weights and recoverable through forensic analysis or extraction attacks \cite{carlini2023extracting}. This aligns with \textit{Micro Star v. FormGen Inc.} \cite{microstar1998}, where the Ninth Circuit held that numerical files were infringing copies because they served as machine-readable instructions to recreate protected works. Similarly, the presence of a secondary filter that blocks the final view does not erase the fact that the weights themselves are capable of reproduction with the aid of the underlying machine.

\section{Beyond Copyright: The Inescapable Web of Contract and Tort Liability}

Even if certain forms of training were permissible under the copyright law, developers may still face substantial exposure under \emph{contract} and \emph{tort/unfair competition} doctrines. Explicit licenses and website terms can restrict access and downstream use regardless of copyright exceptions, and unauthorized large-scale appropriation of curated resources may trigger tort or unfair-competition liability. We show that these independent layers of contract and tort liability can constrain training even when copyright defenses might apply, making copyright-based defenses alone insufficient.

\subsection{Breach of Explicit License Agreements (Signed/Viral Licenses)}
A distinct but equally critical vector of liability arises from contract law. When developers ingest data subject to explicit license agreements—such as non-commercial use only or academic license clauses—they bind themselves to specific contractual obligations. Unlike copyright liability, which depends on statutory interpretation, this liability stems from the voluntary breach of a binding promise.

Courts have long upheld that private contracts can impose stricter limitations on data usage than copyright law itself. The seminal case of \textit{ProCD, Inc. v. Zeidenberg} \cite{procdzeidenberg} provides the definitive precedent. In this case, the Seventh Circuit enforced a consumer-use only license against a defendant who scraped uncopyrightable telephone listings for commercial resale. Crucially, the court held that even where copyright law does not protect the data, the defendant was legally liable for breach of contract. As a result, the court enforced the license terms and remanded the case for the issuance of a permanent injunction to stop the unauthorized use. In the context of AI, this establishes that a developer who trains a commercial model on research-only data faces strict contractual liability and injunctive relief, regardless of whether the data itself is copyrightable.

\subsection{Breach of Website Terms of Service (ToS / Browsewrap)}
Even without a signed contract, courts have treated enforceable website terms as capable of restricting automated access and bulk copying. This matters because many sites expressly prohibit bots, scraping, or downstream use for model training. Contract law thus enables rights holders to restrict access and use even where copyright defenses might be debatable.

Courts have recognized that private contractual terms can restrict data collection and reuse, effectively overriding the public domain status or statutory exceptions of the underlying material. In \textit{Ryanair v. PR Aviation} \cite{ryanairpr}, the court held that a database owner could enforce contract terms prohibiting scraping, even if the database itself lacked copyright protection. This principle extends even to jurisdictions often cited as safe havens for AI training, such as Japan. While Copyright Act Article 30-4 broadly authorizes information analysis, courts have clarified that this does not immunize developers from contractual liability for unauthorized access. In \textit{Yomiuri Shimbun v. News Aggregator} \cite{yomiurionline}, the Intellectual Property High Court of Japan treated the violation of a site’s prohibition on automated access as an actionable tort. Thus, whether in strict or permissive copyright regimes, valid contractual restrictions can block the initial access required to obtain training data, effectively rendering the copyright defense moot.

\subsection{Tort/Unfair Competition: Liability for Free-Riding on Investment}
Where no explicit contract exists, large-scale appropriation of curated resources can still trigger tort-like liability or unfair competition claims. The legal concern is not merely the copying of isolated facts, but the systematic capture of value created by another party’s costly investment in collecting, organizing, and maintaining a dataset. 

Courts have increasingly recognized that the unauthorized appropriation of such investment constitutes an actionable wrong. In \textit{Tsubasa System v. Toppan} \cite{tsubasasystem} (Japan) and \textit{CV-Online v. Melons} \cite{cvonline} (EU), database scraping was held to be unlawful where it amounted to an unfair free-ride on the plaintiff’s substantial technical and financial efforts. In the AI setting, this doctrine maps cleanly onto the extraction of high-effort corpora—such as professional news archives and specialized legal datasets—whose primary economic value lies in the curation itself. Appropriating this concentrated value to train a commercial model without compensation can therefore be actionable under unfair competition principles, providing a critical layer of protection even when copyright claims remain uncertain.

\section{No Profiting from Misappropriation: Unjust Enrichment and Disgorgement}

The contemporary AI business model often rests on a structural asymmetry: developers can dramatically reduce costs by avoiding licensing or curation expenses, while rights holders receive no compensation. This asymmetry is not only a policy concern but also bears directly on liability and remedies, since unjust enrichment and disgorgement doctrines aim to strip wrongful gains. We therefore examine avoided-cost benefits, head-start advantages, and the limits of post-hoc deletion as a deterrent.

\subsection{Illicit Gains from Training Data: Avoided Costs and Head-Start Advantages}
The data ingestion paradigm is built on a fundamental economic shortcut: using protected content as computational scaffolding to avoid the immense costs of lawful data acquisition or synthesis.

\paragraph{High-Value Signals as Avoided Costs}
The economic gain from infringement is best measured as avoided cost. Data valuation research shows that AI performance does not scale linearly with volume; rather, it depends disproportionately on a small subset of high-quality signals such as professional journalism or curated books that drive reasoning and reliability \cite{ghorbani2019}. Legally, the illicit gain is not the retail price of these works, but the massive expenditure saved by bypassing the market to obtain these essential signals. Under \textit{Sheldon v. Metro-Goldwyn Pictures} \cite{sheldonmgm}, disgorgement requires surrendering the profits to the extent they are attributable to the wrongful shortcut, including avoided costs and any incremental commercial advantage causally linked to the infringement.

\paragraph{The Head Start Doctrine: Deletion Does Not Remove the Lead}
The Head Start doctrine dictates that defendants cannot retain competitive advantages gained through wrongful shortcuts, even if they attempt later technical fixes. In \textit{Integrated Cash Management Services, Inc. v. Digital Transactions, Inc.} \cite{integratedcashmgmt}, the court recognized injunctive relief as appropriate to neutralize the ``head start'' attributable to improper use of trade secrets. Likewise, in AI, unlawful corpora materially accelerate model convergence and time-to-market. Because these data points have already performed their functional role in shaping the model's maturity, the developer continues to enjoy a persistent acceleration in the market. Deleting the source files post-training does not neutralize this structural advantage, as the intelligence gained remains in the model's parameters.

\subsection{Moral Hazard: Why Delete if Caught Is Not an Adequate Deterrent}
An enforcement regime that allows companies to keep the principal benefit of wrongful acquisition after paying a manageable penalty risks creating moral hazard. If the expected downside is lower than the cost of licensing, misappropriation becomes a rational strategy. This requires structural relief that extends to the model itself, as seen in precedents where regulators mandated algorithmic disgorgement to prevent firms from retaining the proceeds of unlawful ingestion (\textit{e.g., In re Everalbum} \cite{ftceveralbum}).

\paragraph{The Fallacy of Efficient Breach}
If the remedy for wrongful data acquisition is limited to deleting the source files after the model has shipped, the law effectively subsidizes the shortcut. A simple deterrence intuition suggests that a rule requiring only the return of an item if caught invites systematic misconduct. The AI analog is clear: if firms can retain the capability gained from restricted data after deleting the raw dataset, they have effectively laundered the benefit of the violation into a permanent commercial asset. This problem is compounded by the technical limits of delete the data assurances. Once high-value data has been used to train a large model, its influence is distributed throughout a high-dimensional parameter space. Accordingly, a promise to delete the dataset may not meaningfully remove the competitive advantage created by the earlier use, especially where the same training pipeline and derivative checkpoints persist. Without reaching the model itself, the law fails to address the persistent residual advantages of the initial illegality.

\section{Alternative Views}
We acknowledge that the legal landscape regarding AI training remains unsettled, and forceful arguments exist against imposing strict liability. Below, we address three primary counter-arguments often raised by the ML community. While these positions raise valid concerns about innovation and feasibility, we contend that existing legal doctrines ultimately favor a framework of accountability for data ingestion.

\subsection{The Dataset Creation as Fair Use Argument}
\textbf{The View:} Proponents argue that our premise is untenable because training a model is functionally equivalent to human learning—a non-infringing act. Relying on \textit{Authors Guild v. Google} \cite{authorsgoogle}, they contend that creating a training dataset constitutes transformative fair use because the model analyzes statistical patterns rather than engaging in the expressive use of the original works.

\textbf{Our Response:} This reliance on the \textit{Google Books} precedent is misplaced as it ignores the fundamental shift in purpose and market harm.
\begin{itemize}
    \item \textbf{From Search to Substitution:} In \textit{Google Books}, the copying was deemed fair because the resulting search index served as a complement to the originals, directing users to them. In contrast, Generative AI datasets are used to create market substitutes. As established in \textit{Thomson Reuters v. Ross} \cite{thomsonross}, copying to build a product that competes with the source material's value is not fair use.
    \item \textbf{The Ingestion Market:} Unlike the search era, a market now exists for licensing data for AI training. By bypassing this market to create datasets for free, developers usurp the copyright holder's right to license their work for computational analysis—a harm recognized in \textit{American Geophysical Union v. Texaco} \cite{texaco}.
\end{itemize}

\subsection{The Data Ingestion as Fair Use Argument}
\textbf{The View:} Proponents argue that our premise is also untenable since data ingestion is an intermediate step analogous to the reverse engineering in \textit{Sega v. Accolade} \cite{segaaccolade}, which was found as a fair use. They contend that intermediate copying is permissible if it is the only way to access unprotected functional elements for legitimate analysis.

\textbf{Our Response:} This argument fails because it misapplies the narrow exception in \textit{Sega}, which strictly requires a non-competing end use.
\begin{itemize}
    \item \textbf{The Condition of Complementarity:} The \textit{Sega} ruling was not a blanket permission. It excused intermediate copying only because the resulting end-products (compatible games) were complementary and did not substitute for the original work. The exception is contingent on the creation of a non-infringing, non-competing asset.
    \item \textbf{Market Substitution:} Generative AI violates this critical condition. Unlike the complementary goods in \textit{Sega}, AI models function as direct market substitutes for their training data. As established in \textit{Thomson Reuters v. Ross} \cite{thomsonross}, the intermediate copying defense is unavailable when the analysis is used to generate products that compete with the original source. This substitution breaks \textit{Sega}’s required nexus, making the ingestion infringing.
\end{itemize}

\subsection{The Public Interest and Transaction Costs Argument}
\textbf{The View:} Finally, it is argued that the transaction costs of licensing billions of documents are prohibitively high. Imposing liability could stifle innovation and bankrupt AI labs, contrary to the public interest. Therefore, copyright exceptions should be expanded to prevent market failure.

\textbf{Our Response:} While transaction costs are real, they do not justify a blanket exemption from liability.
\begin{itemize}
    \item \textbf{Licensing Solutions:} High transaction costs are a market design problem, not a legal justification for using data without permission. In practice, collective licensing mechanisms (e.g., the Copyright Clearance Center) exist to reduce this friction. Difficulty in clearing rights does not make unauthorized ingestion legal.

    \item \textbf{Long-term Sustainability:} An AI supply chain built on uncompensated extraction is fragile. As prior work warns about model collapse~\cite{shumailov2023curse}, training successive generations of models on data that increasingly include model-generated content can lead to a degradation in the quality unless access to original, human-generated data is preserved. Supporting creators' incentives is therefore aligned with the long-term health of the AI ecosystem.
\end{itemize}

\section{Call to Action: From Post-hoc Mitigation to Ex-ante Process Compliance}

Our analysis confirms that legal liabilities are established at the exact moment of data acquisition or ingestion. Consequently, post-hoc mitigation techniques (like guardrails and unlearning) are legally insufficient. The industry must undergo a fundamental paradigm shift from cleaning up after the fact to getting it right from the start: a transition to \emph{Ex-ante Process Compliance}. While the goal is to establish a verifiable chain of title before training begins, the responsibilities are shared between those who build the systems and those who set the rules.

\subsection{For Engineers (Researchers \& Developers): Engineering Verifiability}
The academic and development communities must abandon the black box approach. Instead of relying on trust or probabilistic unlearning, they must design architectures that allow for mechanical verification of data hygiene.

\begin{itemize}
    \item \textbf{Cryptographic Transparency (The Glass Pipeline):}
    To balance secrecy with accountability, developers can build a ``Glass Pipeline'' using cryptographic tooling. Using standards such as C2PA~\cite{c2pa}, each weight update can be tied to verifiable provenance credentials from rights holders (positive verification). Complementarily, Zero-Knowledge Proofs (ZKPs) can enable auditors to verify that restricted works were excluded from training without revealing the full corpus (negative verification). Together, these mechanisms replace trust-based assurances with auditable compliance claims.

    \item \textbf{Architectural Reversibility (Git for Models):}
    Recognizing that knowledge from training data cannot always be cleanly removed once it has been integrated into model parameters, the system should support reliable rollback. Researchers should implement high-frequency checkpointing, which functions as version control for model weights, so that training can be replayed from a known-good state. If a source $S$ is later determined to be unauthorized, the developer should avoid attempting to subtract it from the final model via unlearning. Instead, they should perform a surgical re-branching by reverting to the checkpoint $\theta_{t < \mathrm{ingestion}(S)}$ from before $S$ was ingested and then retraining forward on a clean data path. This approach provides a concrete and auditable way to demonstrate that the model was produced without relying on the contested source.

\end{itemize}

\subsection{For Policymakers: Structuring the Compliance Market}
Technical solutions are futile if the legal framework makes compliance prohibitively expensive. Policymakers must reform the market to ensure that obeying the law is cheaper and easier than stealing data.

\begin{itemize}
    \item \textbf{Centralized Licensing via Data Trusts:} 
    Navigating rights negotiations with millions of individual creators is transactionally impossible (the Tragedy of the Anticommons). To solve this, regulators should establish Data Trusts or Collective Management Organizations. 
    \par
    These entities would act as centralized clearinghouses, offering licenses on FRAND (Fair, Reasonable, and Non-Discriminatory) terms. Developers would pay a single standardized fee to the Trust, which then algorithmically distributes royalties to authors—potentially weighted by metrics like Data Shapley values. This eliminates the excuse that licensing is too hard while ensuring creators get paid.

    \item \textbf{Mandatory Rights Declaration (The Binary Beacon):} 
    The current reliance on voluntary opt-outs (like \texttt{robots.txt}) is legally ambiguous and unfair to creators. We propose a statutory Binary Beacon regime: a mandatory, machine-readable signal that explicitly declares AI training status.
    \par
    Crucially, legislation must shift the default presumption from Opt-Out to Opt-In. Any content without a clear beacon should default to permission denied. This shifts the burden of verification away from the artist—who currently has to chase down scrapers—and places strict liability on the data acquirer to ensure they have explicit permission before ingestion.
\end{itemize}

\section{Conclusion}

\textbf{This paper has argued that post-hoc mitigation strategies do not, as a matter of law, eliminate liability arising from the training process.} Liability is not determined solely by a model’s output; rather, infringement is established by the underlying acts of acquisition and reproduction. Once these unlawful acts are completed, technical adjustments to outputs cannot retroactively cure the illegality. This exposes developers to severe risks, including massive financial penalties and algorithmic disgorgement—the destruction of the model itself.

Our analysis confirms this through three key findings. First, unauthorized ingestion constitutes an incurable \emph{Completed Act} ($t=0$), and model weights function as \emph{Fixed Copies} that ontologically preserve infringing value, rendering post-hoc filters legally irrelevant. Second, a \emph{Global Encirclement} of contract and tort liabilities effectively overrides standard copyright defenses regardless of fair use claims. Third, the doctrine of \emph{Unjust Enrichment} mandates the disgorgement of any algorithmic value derived from misappropriated foundations, denying the retention of illicitly gained benefits.

To avoid the risks, the industry must shift fundamentally toward \emph{Ex-ante Process Compliance}. Technologists must abandon opaque practices for architectures ensuring verifiable lineage, while policymakers must reconstruct markets to facilitate standardized licensing and clear consent. The focus must move from retroactively patching leaks to proactively securing the source.

Ultimately, the legitimacy of an AI system depends on the integrity of its inputs, not the safety of its outputs. True compliance is not about what a model \emph{says}, but about how it came to \emph{know}.

\section*{Acknowledgements}
This work is partially supported by JST CREST JPMJCR21D1, JST BOOST JPMJBY24A6, and JSPS KAKENHI JP23K16940. 

\bibliography{example_paper}

@inproceedings{brown2020language,
 author = {Brown, Tom and Mann, Benjamin and Ryder, Nick and Subbiah, Melanie and Kaplan, Jared D and Dhariwal, Prafulla and Neelakantan, Arvind and Shyam, Pranav and Sastry, Girish and Askell, Amanda and Agarwal, Sandhini and Herbert-Voss, Ariel and Krueger, Gretchen and Henighan, Tom and Child, Rewon and Ramesh, Aditya and Ziegler, Daniel and Wu, Jeffrey and Winter, Clemens and Hesse, Chris and Chen, Mark and Sigler, Eric and Litwin, Mateusz and Gray, Scott and Chess, Benjamin and Clark, Jack and Berner, Christopher and McCandlish, Sam and Radford, Alec and Sutskever, Ilya and Amodei, Dario},
 booktitle = {Advances in Neural Information Processing Systems},
 editor = {H. Larochelle and M. Ranzato and R. Hadsell and M.F. Balcan and H. Lin},
 pages = {1877--1901},
 publisher = {Curran Associates, Inc.},
 title = {Language Models are Few-Shot Learners},
 volume = {33},
 year = {2020}
}

@article{bommasani2021opportunities,
  title={On the Opportunities and Risks of Foundation Models},
  author={Rishi Bommasani and Drew A. Hudson and Ehsan Adeli and Russ Altman and Simran Arora and Sydney von Arx and Michael S. Bernstein and Jeannette Bohg and Antoine Bosselut and Emma Brunskill and Erik Brynjolfsson and Shyamal Buch and Dallas Card and Rodrigo Castellon and Niladri Chatterji and Annie Chen and Kathleen Creel and Jared Quincy Davis and Dora Demszky and Chris Donahue and Moussa Doumbouya and Esin Durmus and Stefano Ermon and John Etchemendy and Kawin Ethayarajh and Li Fei-Fei and Chelsea Finn and Trevor Gale and Lauren Gillespie and Karan Goel and Noah Goodman and Shelby Grossman and Neel Guha and Tatsunori Hashimoto and Peter Henderson and John Hewitt and Daniel E. Ho and Jenny Hong and Kyle Hsu and Jing Huang and Thomas Icard and Saahil Jain and Dan Jurafsky and Pratyusha Kalluri and Siddharth Karamcheti and Geoff Keeling and Fereshte Khani and Omar Khattab and Pang Wei Koh and Mark Krass and Ranjay Krishna and Rohith Kuditipudi and Ananya Kumar and Faisal Ladhak and Mina Lee and Tony Lee and Jure Leskovec and Isabelle Levent and Xiang Lisa Li and Xuechen Li and Tengyu Ma and Ali Malik and Christopher D. Manning and Suvir Mirchandani and Eric Mitchell and Zanele Munyikwa and Suraj Nair and Avanika Narayan and Deepak Narayanan and Ben Newman and Allen Nie and Juan Carlos Niebles and Hamed Nilforoshan and Julian Nyarko and Giray Ogut and Laurel Orr and Isabel Papadimitriou and Joon Sung Park and Chris Piech and Eva Portelance and Christopher Potts and Aditi Raghunathan and Rob Reich and Hongyu Ren and Frieda Rong and Yusuf Roohani and Camilo Ruiz and Jack Ryan and Christopher Ré and Dorsa Sadigh and Shiori Sagawa and Keshav Santhanam and Andy Shih and Krishnan Srinivasan and Alex Tamkin and Rohan Taori and Armin W. Thomas and Florian Tramèr and Rose E. Wang and William Wang and Bohan Wu and Jiajun Wu and Yuhuai Wu and Sang Michael Xie and Michihiro Yasunaga and Jiaxuan You and Matei Zaharia and Michael Zhang and Tianyi Zhang and Xikun Zhang and Yuhui Zhang and Lucia Zheng and Kaitlyn Zhou and Percy Liang},
  journal={arXiv preprint arXiv:2108.07258},
  year={2021}
}

@inproceedings{bourtoule2021machine,
  title={Machine unlearning},
  author={Bourtoule, Lucas and Chandrasekaran, Varun and Choquette-Choo, Christopher A and Jia, Hengrui and Travers, Adelin and Zhang, Baiwu and Lie, David and Papernot, Nicolas},
  booktitle={2021 IEEE symposium on security and privacy (SP)},
  pages={141--159},
  year={2021},
  organization={IEEE}
}

@article{eldan2023who,
  title={Who's Harry Potter? Approximate Unlearning in LLMs},
  author={Eldan, Ronen and Russinovich, Mark},
  journal={arXiv preprint arXiv:2310.02238},
  year={2023}
}

@inproceedings{meng2022locatingeditingfactualassociations,
    author = {Meng, Kevin and Bau, David and Andonian, Alex and Belinkov, Yonatan},
    title = {Locating and editing factual associations in GPT},
    year = {2022},
    booktitle = {Proceedings of the 36th International Conference on Neural Information Processing Systems},
    articleno = {1262},
    numpages = {14},
    location = {New Orleans, LA, USA},
    series = {NIPS '22}
}

@inproceedings{jang2023knowledge,
  title={Knowledge Unlearning for Mitigating Privacy Risks in Language Models},
  author={Jang, Joel and Yoon, Dongkeun and Yang, Seorie and Cha, Sung Ju and Lee, Moontae and Logeswaran, Lajanugen and Seo, Minjoon},
  booktitle={Proceedings of the 61st Annual Meeting of the Association for Computational Linguistics (Volume 1: Long Papers)},
  pages={14389--14408},
  year={2023},
}

@article{inan2023llama,
  title={Llama guard: Llm-based input-output safeguard for human-ai conversations},
  author={Inan, Hakan and Upasani, Kartikeya and Chi, Jianfeng and Rungta, Rashi and Iyer, Krithika and Mao, Yuning and Tontchev, Michael and Hu, Qing and Fuller, Brian and Testuggine, Davide and others},
  journal={arXiv preprint arXiv:2312.06674},
  year={2023}
}

@article{shumailov2023curse,
  title={The curse of recursion: Training on generated data makes models forget},
  author={Shumailov, Ilia and Shumaylov, Zakhar and Zhao, Yiren and Gal, Yarin and Papernot, Nicolas and Anderson, Ross},
  journal={arXiv preprint arXiv:2305.17493},
  year={2023}
}

@article{bai2022constitutional,
  title={Constitutional AI: Harmlessness from AI Feedback},
  author={Bai, Yuntao and Kadavath, Saurav and Kundu, Sandipan and Askell, Amanda and Kernion, Jackson and Jones, Andy and Chen, Anna and Goldie, Anna and Mirhoseini, Azalia and others},
  journal={arXiv preprint arXiv:2212.08073},
  year={2022}
}

@inproceedings{ghorbani2019,
  title = {Data Shapley: Equitable Valuation of Data for Machine Learning},
  author = {Ghorbani, Amirata and Zou, James},
  booktitle = {Proceedings of the 36th International Conference on Machine Learning (ICML)},
  volume = {97},
  pages = {2242--2251},
  year = {2019},
  publisher = {PMLR}
}

@inproceedings{carlini2023extracting,
    author = {Carlini, Nicholas and Hayes, Jamie and Nasr, Milad and Jagielski, Matthew and Sehwag, Vikash and Tram\`{e}r, Florian and Balle, Borja and Ippolito, Daphne and Wallace, Eric},
    title = {Extracting training data from diffusion models},
    year = {2023},
    isbn = {978-1-939133-37-3},
    publisher = {USENIX Association},
    booktitle = {Proceedings of the 32nd USENIX Conference on Security Symposium},
    articleno = {294},
    numpages = {18},
}

@article{lemley2021,
  title={Fair Learning},
  author={Lemley, Mark A and Casey, Bryan},
  journal={Texas Law Review},
  volume={99},
  year={2021}
}

@article{sag2018,
  title={The new legal landscape for text mining and machine learning},
  author={Sag, Matthew},
  journal={J. Copyright Soc'y USA},
  volume={66},
  pages={291},
  year={2018},
  publisher={HeinOnline}
}

@misc{c2pa,
  title={C2PA Technical Specification},
  author={{Coalition for Content Provenance and Authenticity}},
  year={2023}
}

@article{quang2021,
  title={Does Training AI Violate Copyright Law?},
  author={Quang, Jenny},
  journal={Berkeley Technology Law Journal},
  volume={36},
  number={4},
  pages={1407--1436},
  year={2021}
}

@article{murray2025,
  title={{AI} Training is Fair Use: The Beginning of the End of the Copyright Assault on Gen AI},
  author={Murray, Michael D.},
  journal={SSRN Working Paper},
  year={2025}
}

@article{christensen2021,
  title={A European solution for Text and Data Mining in the development of creative Artificial Intelligence: With a specific focus on articles 3 and 4 of the Digital Single Market Directive},
  author={Christensen, Kristina},
  journal={Stockholm Intellectual Property Law Review},
  volume={4},
  number={2},
  pages={18--33},
  year={2021}
}

@article{lobling2024,
  title={Navigating the Legal Landscape: Technical Implementation of Copyright Reservations for Text and Data Mining in the Era of AI Language Models},
  author={L{\"o}bling, Lisa and Handschigl, Christian and Hofmann, Kai and Schwedhelm, Jan},
  journal={JIPITEC -- Journal of Intellectual Property, Information Technology and E-Commerce Law},
  volume={14},
  number={4},
  year={2024},
  note={Published 29 Feb 2024}
}

@article{dermawan2024,
  title={Text and data mining exceptions in the development of generative AI models: What the EU member states could learn from the Japanese “nonenjoyment” purposes?},
  author={Dermawan, Artha},
  journal={The Journal of World Intellectual Property},
  volume={27},
  number={1},
  pages={44--68},
  year={2024},
  publisher={Wiley Online Library}
}

@article{lemley2023,
  title={How generative AI turns copyright upside down},
  author={Lemley, Mark A},
  journal={Available at SSRN 4517702},
  year={2023}
}

@article{sobel2024,
  title={Elements of style: copyright, similarity, and generative AI},
  author={Sobel, Benjamin},
  journal={Harvard Journal of Law \& Technology},
  volume={38},
  year={2024}
}

@article{rosati2024,
  title={Infringing AI: Liability for AI-generated outputs under international, EU, and UK copyright law},
  author={Rosati, Eleonora},
  journal={European Journal of Risk Regulation},
  pages={1--25},
  year={2024},
  publisher={Cambridge University Press}
}

@article{dornis2025,
  title={Generative AI and the Scope of EU Copyright Law: A Doctrinal Analysis in Light of C-250/25},
  author={Dornis, Tim W. and Lucchi, Nicola},
  journal={International Review of Intellectual Property and Competition Law},
  volume={56},
  number={10},
  pages={1123--1148},
  year={2025}
}

@article{zhang2025,
  title={Input out, output in: towards positive-sum solutions to AI-copyright tensions},
  author={Zhang, Jiawei},
  journal={Journal of Intellectual Property Law \& Practice},
  volume={20},
  number={9},
  pages={594--604},
  year={2025},
  publisher={Oxford University Press}
}

@article{murray2023,
  title={Generative ai art: Copyright infringement and fair use},
  author={Murray, Michael D},
  journal={SMU Sci. \& Tech. L. Rev.},
  volume={26},
  pages={259},
  year={2023},
  publisher={HeinOnline}
}

@misc{marino2025,
      title={Position: Bridge the Gaps between Machine Unlearning and AI Regulation}, 
      author={Bill Marino and Meghdad Kurmanji and Nicholas D. Lane},
      year={2025},
      eprint={2502.12430},
      archivePrefix={arXiv},
      primaryClass={cs.LG},
      url={https://arxiv.org/abs/2502.12430}, 
}

@misc{bartzorder,
  title={Order on Fair Use, Bartz et al. v. Anthropic PBC},
  howpublished={No. C 24-05417 WHA (N.D. Cal. June 23, 2025)},
  year={2025}
}

@misc{thomsonross,
  title={Thomson Reuters Enter. Ctr. GmbH v. Ross Intelligence Inc.},
  howpublished={No. 1:20-cv-00613 (D. Del. Sept. 25, 2023)},
  year={2023}
}

@misc{warholgoldsmith,
  title={Andy Warhol Found. for the Visual Arts, Inc. v. Goldsmith},
  howpublished={598 U.S. 508},
  year={2023}
}

@misc{ftceveralbum,
  title={In the Matter of Everalbum, Inc.},
  howpublished={FTC File No. 1923172},
  year={2021}
}

@misc{procdzeidenberg,
  title={ProCD, Inc. v. Zeidenberg},
  howpublished={86 F.3d 1447 (7th Cir. 1996)},
  year={1996}
}

@misc{authorsgoogle,
  title={Authors Guild v. Google, Inc.},
  howpublished={804 F.3d 202 (2d Cir. 2015)},
  year={2015}
}

@misc{segaaccolade,
  title={Sega Enterprises Ltd. v. Accolade, Inc.},
  howpublished={977 F.2d 1510 (9th Cir. 1992)},
  year={1992}
}

@misc{nytopenai,
  title={The New York Times Co. v. Microsoft Corp. et al.},
  howpublished={No. 1:23-cv-11195 (S.D.N.Y. filed Dec. 27, 2023)},
  year={2023}
}

@misc{gettystability,
  title={Getty Images (US), Inc. v. Stability AI, Ltd.},
  howpublished={No. 1:23-cv-00135 (D. Del. filed Feb. 3, 2023)},
  year={2023}
}

@misc{mai,
  title={MAI Systems Corp. v. Peak Computer, Inc.},
  howpublished={991 F.2d 511 (9th Cir. 1993)},
  year={1993}
}

@misc{redigi,
  title={Capitol Records, LLC v. ReDigi Inc.},
  howpublished={910 F.3d 649 (2d Cir. 2018)},
  year={2018}
}

@misc{microstar1998,
  title = {Micro Star v. FormGen Inc.},
  howpublished = {154 F.3d 1107 (United States Court of Appeals, Ninth Circuit)},
  year = {1998},
}

@misc{texaco,
  title  = {American Geophysical Union v. Texaco Inc.},
  volume = {60},
  reporter = {F.3d},
  pages  = {913},
  court  = {2d Cir.},
  year   = {1994}
}

@misc{sheldonmgm,
  title={Sheldon v. Metro-Goldwyn Pictures Corp.},
  howpublished={309 U.S. 390},
  year={1940}
}

@misc{integratedcashmgmt,
  title={Integrated Cash Management Services, Inc. v. Digital Transactions, Inc.},
  year={1990},
  howpublished={920 F.2d 171 (2d Cir. 1990)},
}

@misc{ryanairpr,
  title={Ryanair Ltd v PR Aviation BV},
  howpublished={Case C-30/14 (CJEU 2015)},
  year={2015}
}

@misc{cvonline,
  title={CV-Online Latvia v Melons},
  howpublished={Case C-762/19 (CJEU 2021)},
  year={2021}
}

@misc{tsubasasystem,
  title={Tsubasa System Co. Ltd. v. Toppan Printing Co. Ltd.},
  howpublished={1780 Hanrei Jiho 25 (Tokyo Dist. Ct. 2001)},
  year={2001}
}

@misc{yomiurionline,
  title={Yomiuri Shimbun v. News Aggregator},
  author={{Intellectual Property High Court of Japan}},
  year={2005},
  howpublished={Judicial decision},
}

@misc{laionkneschke,
  title={Robert Kneschke v. LAION e.V.},
  author={{Hamburg Regional Court (Landgericht Hamburg)}},
  year={2024},
  howpublished={Case No. 310 O 227/23},
  note={Judgment of September 27, 2024}
}
\bibliographystyle{icml2026}

\newpage
\appendix
\onecolumn

\section{Engineering Output Safety vs. Legal Reproduction Liability}

\subsection{The Engineering Perspective: Output Equivalence and Functional Restoration}

From an engineering standpoint, a model can be viewed as a function $f(x)$ that maps inputs to outputs. Compliance, on this view, is achieved by ensuring a safe output distribution. The dominant engineering logic is that even if a model $M$ has been trained on harmful data $D_{\text{harmful}}$, the concern is resolved if post-hoc techniques—such as unlearning, fine-tuning, or guardrails—can adjust the model’s behavior such that it becomes statistically indistinguishable from a counterfactual model $M'$ trained without $D_{\text{harmful}}$. In this framework, the emphasis is placed on present functional safety rather than the historical lineage of how the model was trained.

\subsection{The Legal Perspective: Liability Becomes Fixed at the Moment of Reproduction}

From a legal standpoint, by contrast, liability may arise not only from outputs but also from the act of reproduction itself during training. In many jurisdictions, once data is copied or stored—whether on disk or in memory—for training without authorization, the relevant infringement (or other violation) may already be considered a completed act.

Accordingly, post-hoc interventions such as unlearning or output guardrails may reduce the risk of future harmful outputs, but they function only as prospective risk-mitigation measures. They do not retroactively cure the unlawfulness of the earlier completed act of unauthorized copying or use. In other words, legal liability can become fixed at the point of reproduction, and post-hoc mitigation cannot, by itself, operate as a legal exculpation.

\section{Technical Taxonomy of Post-Hoc Mitigation Methods}

In this paper, we define the following two types of post-hoc mitigation techniques: inference-time control and parameter modification. 

\subsection{Inference-Time Control (Prompt/Output Level)}
This category includes approaches that constrain model behavior without changing the model’s internal parameters. Because these measures are relatively inexpensive to deploy and can be updated quickly without retraining, they have become the dominant standard in commercial AI systems.

Safety filters operate as an external wrapper around a trained, frozen model. Systems such as Llama Guard~\cite{inan2023llama} or OpenAI’s moderation endpoints typically rely on separate, lightweight classifiers that monitor the interaction in real time. These components can detect and block prompts that request copyrighted material, and can also suppress, truncate, or rewrite outputs that appear to closely resemble protected works. Functionally, these techniques restrict what the user is allowed to see or receive, while leaving the underlying model weights—and any latent knowledge encoded in them—unchanged.

\subsection{Model Modification (Parameter Level)}
This category aims to alter the model’s internal state so that specific knowledge or capabilities are removed (or at least substantially weakened) within the weights themselves. 
Although academic interest in these methods is substantial, practical deployment remains difficult, in part because targeted edits can affect model stability and general performance.

\textbf{Machine Unlearning:}
Machine unlearning methods seek to transform a model trained on dataset $D$ into one that approximates a counterfactual model trained on $D \setminus D_{\text{forget}}$ (i.e., the original dataset with a specified forget set removed).
\begin{itemize}
    \item \textit{Exact unlearning:} Methods such as SISA (Sharded, Isolated, Sliced, and Aggregated)~\citep{bourtoule2021machine} can provide strong guarantees by retraining only the affected sub-models. However, at the scale of large language models, this approach is often computationally prohibitive.
    \item \textit{Approximate unlearning:} As a result, most practical work focuses on approximations, such as targeted \textit{gradient ascent} updates that attempt to undo the learning signal by reducing the likelihood of specific tokens or associations~\citep{jang2023knowledge}.
\end{itemize}
\textbf{Model Editing:}
Model editing techniques attempt to locate where particular factual associations are represented in the network and then modify those representations directly~\citep{meng2022locatingeditingfactualassociations}. 

\textbf{Steering and Constitutional AI:}
Approaches like \textit{Constitutional AI}~\cite{bai2022constitutional} train models to critique and revise their own outputs according to articulated safety principles. The result is an internalized refusal mechanism: the model may decline to generate infringing content in response to certain requests, even though the relevant information may still remain represented within the parameters.

Unlike inference-time guardrails, which restrict behavior at the inference, these methods aim to overwrite information within the weight matrices themselves, analogous to performing precise surgery to alter a specific memory.


\end{document}